\documentclass[12pt]{iopart}

\usepackage{graphicx}

\usepackage{iopams}  
\usepackage{amssymb}
\usepackage{bm}
\usepackage{siunitx}
\newcommand{\AVE}[1]{\ensuremath{\langle {#1} \rangle}}

\newcommand{\dpone}[2]{\ensuremath{\displaystyle\frac{\partial {#1}}{\partial {#2}}}}

\newcommand{\bF}{\ensuremath{\bm{F}}}

\newcommand{\bI}{\ensuremath{\bm{I}}}

\newcommand{\bM}{\ensuremath{\bm{M}}}

\newcommand{\bT}{\ensuremath{\bm{T}}}

\newcommand{\bX}{\ensuremath{\bm{X}}}

\newcommand{\be}{\ensuremath{\bm{e}}}
\newcommand{\bff}{\ensuremath{\bm{f}}}

\newcommand{\bp}{\ensuremath{\bm{p}}}
\newcommand{\bq}{\ensuremath{\bm{q}}}
\newcommand{\bbr}{\ensuremath{\bm{r}}}

\newcommand{\bx}{\ensuremath{\bm{x}}}

\begin{document}

\title
[ 
Contraction in active microtubule networks
]
{Connecting macroscopic dynamics with microscopic properties in active microtubule network contraction}

\author{Peter J. Foster$^{1,5}$, Wen Yan$^{2,5}$, Sebastian F\"{u}rthauer$^{2}$, Michael J. Shelley$^{2,3}$, Daniel J. Needleman$^{1,4}$}
\address{$^1$John A. Paulson School of Engineering and Applied Sciences, FAS Center for Systems Biology, Harvard University, Cambridge MA, United States}
\address{$^2$Center for Computational Biology, Flatiron Institute, Simons Foundation, New York, NY, United States}
\address{$^3$Courant Institute of Mathematical Science, New York University, New York, NY, United States}
\address{$^4$Department of Molecular and Cellular Biology, Harvard University, Cambridge MA, United States}
\address{$^5$Equal contribution}

\ead{\mailto{peterfoster@fas.harvard.edu},\mailto{ wyan@flatironinstitute.org}}
\vspace{10pt}
\begin{indented}
\item[]June 2017
\end{indented}

\begin{abstract}
The cellular cytoskeleton is an active material, driven out of equilibrium by molecular motor proteins. It is not understood how the collective behaviors of cytoskeletal networks emerge from the properties of the network's constituent motor proteins and filaments. Here we present experimental results on networks of stabilized microtubules in \emph{Xenopus} oocyte extracts, which undergo spontaneous bulk contraction driven by the motor protein dynein, and investigate the effects of varying the initial microtubule density and length distribution. We find that networks contract to a similar final density, irrespective of the length of microtubules or their initial density, but that the contraction timescale varies with the average microtubule length. To gain insight into why this microscopic property influences the macroscopic network contraction time, we developed simulations where microtubules and motors are explicitly represented. The simulations qualitatively recapitulate the variation of contraction timescale with microtubule length, and allowed stress contributions from different sources to be estimated and decoupled.
 
\end{abstract}

%
\vspace{2pc}
\noindent{\it Keywords}: active matter, microtubule, motor protein, dynein, \emph{Xenopus} extract
%
%
%
%

\section{Introduction}
Active matter is a class of materials held out of equilibrium by the local
conversion of energy from a reservoir into mechanical work at the scale of the
system's components \cite{Marchetti:2013bp}. Active matter can exhibit 
emergent behaviors, such as collective motion and pattern formation, on lengthscales much larger than the size of the system's constituents.

Here we consider cytoskeletal networks,
 which are living, active systems \cite{Prost:2015ev} composed of polar polymeric filaments and held out of equilibrium by molecular motor proteins which convert chemical energy into
mechanical work. Cytoskeletal networks are responsible for a number of cell biological processes, including cell division and chromosome segregation. The active behaviors of cytoskeletal networks have been experimentally investigated in both simplified purified mixtures \cite{Sanchez:2012gt,Linsmeier:2016ff,Surrey:2001cz,Surrey:1997p7217} and in complex systems in cells and cell  extracts \cite{Foster:2015gfa,Brugues:2014bp,Naganathan:2014fc,Mayer:2010kt}. However, it remains poorly understood how the dynamics and architecture of these networks are shaped
by the properties of their constituent filaments and motors.

The past several years have seen several studies of the dynamics and
mechanics of microtubule (MTs) / motor protein suspensions (see \cite{shelley_dynamics_2016} for a
review).  A predominant thread of experiments have focused on dense,
nematically aligned phases of MT bundles and immersed layers which
show dynamics driven by extension of material along the direction
of orientational order \cite{sanchez}. It is believed that this
extensile motion is driven by the polarity sorting of anti-aligned MTs
by multimeric kinesin motors. Theoretical modeling of the material
stresses produced by polarity sorting indeed yields active stress tensors
that are anisotropic, and arise from ensemble averages of terms of the form
$\alpha \bp \bp$, where $\bp$ is an MT orientation vector, and $\alpha<0$
gives extensile flows along the $\bp$ direction \cite{gao_multiscale_2015}. 
Different dynamics can follow from different MT interactions, and clustering of MT ends is a generic mechanism leading to contraction \cite{nedelec_biorxic}. This has been previously explored in other purified systems containing kinesin motor proteins \cite{Nedelec:2001kq,Hentrich:2010jh}. In this work, as in our earlier work of stabilized MTs in meiotically-arrested \emph{Xenopus} oocyte extracts \cite{Foster:2015gfa} and in several related studies \cite{rick_biorxiv,White,Taunenbaum}, we consider a spontaneously formed MT network whose dynamics
is driven by minus-end bundling by dynein motors. This leads to an
apparently isotropic contraction, and we proposed a continuum model \cite{Foster:2015gfa}, which quantitatively reproduced the experimental results. Crucially, the model helped explain why networks contract to a preferred final density regardless of sample geometry and motor concentration. 
 The continuum model assumed that the
active material stress tensors were isotropic and of the form $f(\rho)\bI$,
with $f<0$ a combination of contractile motor stresses and repulsive
steric stresses. 

Here, we explore the same system as in our previous work \cite{Foster:2015gfa} further by varying the initial MT density and the length distribution of MTs and measuring the resulting effects on the contraction process. Consistent with the predictions of our active fluid model, we find that the final network density varies little across a wide range of conditions. Intriguingly, we also find a correlation between the timescale of the contraction process and the average length of MTs in the system. To gain insight into how this microscopic parameter influences the macroscopic contraction timescale, we developed simulations of the contraction process. These simulations recapitulate many key aspects of the experimental results, including the variation of contraction timescale with MT length, and allow access to many aspects of the system difficult to measure experimentally, such as the spatial distribution and nature of the material stress tensors. Taken together, these results further support the underlying assumptions and key predictions of our active fluid model, and provide a framework for future studies.

\section{Methods}
\subsection{Preparation of \emph{Xenopus} extracts}
Extracts were prepared from freshly laid \emph{Xenopus} oocytes as described
previously \cite{Anonymous:2006p8371}. Fresh extracts were sequentially filtered
through 2.0, 1.2, and 0.2 $\mu$m filters before being flash frozen in liquid
nitrogen and stored at -80$^{\circ}$C until use.

\subsection{Preparation of microfluidic devices}
Microfluidic device templates were designed using AutoCAD 360 and Silhouette
Studio software. Device templates were cut from 125 $\mu$m tape (3M
Scotchcal) using a Silhouette Cameo die cutter, and adhered to petri dishes to
create a master. PDMS (Sylgard 184, Dow Corning) was mixed at the standard 10:1
ratio, poured onto masters, degassed under vacuum, and cured overnight at
60$^{\circ}$C. The cured devices were then removed from the masters and inlets
and outlets were created using biopsy punches. Devices and coverslips were
treated with air plasma using a corona device, bonded, and loaded with
passivation mix composed of 5 mg/mL BSA and 2.5\% (w/v) Pluronic F-127 before
overnight incubation at 12$^{\circ}$C. 

\subsection{Bulk contraction assay}
The bulk contraction assay was performed as previously described
\cite{Foster:2015gfa}. Briefly, Alexa-647 labeled tubulin was added to 20 $\mu$L
\emph{Xenopus} extract at a final concentration of $\approx$1 $\mu$M. Then, 0.5 $\mu$L
of taxol suspended in DMSO was then added to the extract to the indicated final
concentration. Extracts were then loaded into passivated microfluidic devices,
sealed with vacuum grease, and imaged using spinning disk confocal microscopy
(Nikon TE2000-E microscope, Yokugawa CSU-X1 spinning disk, Hamamatsu ImagEM
camera, 2x objective, Metamorph acquisition software). The time $t=0$ corresponds to when imaging begins, typically $<1$ min after taxol addition. Images were analyzed using ImageJ and custom MATLAB software. The $\epsilon(t)$ curves were fit using time points when $\epsilon(t)>0.1$.

\subsection{Initial and final density estimation}
In order to estimate network densities, we first assume that Alexa-647 labeled tubulin uniformly
incorporates into MTs, and the overall concentration of tubulin is
taken to be constant and equal to the previously measured value of 18 $\mu$M
\cite{Parsons:1997fs}. From this we can take,
$$18 \mu M = \frac{ \sum_{i=1}^N M_i}{\sum_{i=1}^N V_i} = \frac{\sum_{i=1}^N \beta I_i (\ell A)} {\sum_{i=1}^N \ell A} =\beta \langle I\rangle_N,$$
where $M_i$ is the tubulin mass in pixel i,  $\beta$ is a constant conversion factor between concentration in micromolar and fluorescence intensity, $\ell$ is the depth of the sample, $A$ is the area of the pixel, $V_i=\ell A$ is the volume corresponding to pixel i,  $N$ is the total number of pixels, $I_i$ is the measured intensity of pixel i, $A$ is the area of pixel i, and $\langle I\rangle_N$ is the measured intensity averaged over all pixels in the channel. From this we can infer, 
$$\beta = \frac{18 \mu M}{\langle I \rangle_N}.$$
The intensity at each pixel in the network contains a contribution $N_i$ from
polymerized tubulin and a contribution $B_i$ from monomeric, unpolymerized tubulin. The
signal from monomeric tubulin is assumed to be constant and homogeneous
throughout the channel. Thus, at the given time point where the background and
network intensities are measured, the average concentration of polymerized tubulin in
the network is given by, 
\begin{eqnarray}
\AVE{\rho }(t)& =  \frac{\beta }{N_{network}} \sum_{i=1}^{N_{network}}N_i = \frac{\beta }{N_{network}} \sum_{i=1}^{N_{network}}[I_i-B], \nonumber \\
& = \beta [\langle I\rangle_{network} -B]=18 \mu M \frac { \langle I\rangle_{network} -B}{\langle I \rangle_N},\nonumber
\end{eqnarray}
where $N_{network}$ is the number of pixels in the network at the time point,
and $\langle I\rangle_{network} $ is the intensity averaged over all pixels in
the network.

To estimate the network density, the frame at the time closest to $t = T_c + \tau$ was selected, where $\tau$ is the characteristic contraction time and $T_c$  is the lag time between the beginning of imaging and the beginning of network contraction \cite{Foster:2015gfa}. This frame was
 corrected for inhomogeneous illumination and the camera's dark current,
and analyzed as above to find $\AVE{\rho }(T_c + \tau)$. As we assume that no
MTs are created, destroyed, or added to the network during the
contraction process, the total mass of MTs in the network must be
conserved. Thus,
$$M_{network} = \AVE{\rho}_0 V_0 = \AVE{\rho}(t=T_c + \tau) V_{T_c +\tau} = \AVE{\rho}_F V_F,$$
where $V_0,V_{T_c +\tau},V_F$ are the volume of the network at the beginning, at time $T_c +\tau$, and at the final state.
Then,
$$
\AVE{\rho}(0) = \AVE{\rho}(T_c + \tau) \frac {V_{T_c +\tau}}{V_0} =\AVE{\rho}(T_c + \tau) \frac
{W_{T_c +\tau}H_{T_c +\tau}L_{T_c +\tau}}{W_0H_0L_0} ,
$$
where $W,H,L$ are the width, height and length of the network at different times, denoted by the subscripts as with the volume.
We assume that, as the network is pinned at the channel's inlet and outlet,
there is no change in volume along the channel's length, and thus $L_0 =
L_{T_c+\tau}$. Combining Eqns. \ref{eqn:Eq1} and  \ref{eqn:Eq2} in Results, 
$$\epsilon(t=T_c+\tau) = \epsilon_{\infty}(1-e^{-1}) = \frac{W_0 - W(t=T_c +
\tau)}{W_0} = 1-\frac{W_{T_c + \tau}}{W_0},$$
which simplifies to,
$$\frac{W_{T_c + \tau}}{W_0} = 1 -\epsilon_{\infty}(1-e^{-1}). $$
We further assume that the change in the network's height follows the same
functional form as the change in width and thus,
$$
\AVE{\rho}(0) =\AVE{\rho}(T_c + \tau) [1 -\epsilon_{\infty}(1-e^{-1}) ]^2 .
$$
For the final density of the network, similar reasoning can be used to show,
$$
\AVE{\rho}_F = \frac{1}{(1-\epsilon_\infty)^2} \AVE{\rho}_0 = \frac {[1
  -\epsilon_{\infty}(1-e^{-1}) ]^2 }{(1-\epsilon_\infty)^2} \AVE{\rho}(T_c +\tau).
$$

\subsection{Measurement of MT length distributions}
Stabilized MTs were dissociated from motor proteins and fixed as
previously described \cite{Mitchison:2013gt}. MTs were allowed to assemble at room
temperature for 30 minutes (unless otherwise noted) before 5 $\mu$L of extract was
diluted into 50 $\mu$L of MT Dissociation Buffer (250 mM NaCl, 10 mM
K-HEPES, pH 7.7, 1 mM MgCl$_2$, 1 mM EGTA, and 20 $\mu$M taxol). After a 2
minute incubation, 100 $\mu$L of Fix Buffer (0.1\% glutaraldehyde in 60\%
glycerol, 40\% BRB80) was added and incubated for 5 minutes. 1 mL of Dilution
Buffer (60\% glycerol, 40\% BRB80) was added to dilute the sample, and 2
$\mu$L of the diluted sample was spread between a slide and a 22$\times$22
mm$^2$ coverslip. After waiting 30 minutes to allow the MTs to adhere
to the coverslip, MTs were imaged using spinning disk confocal
microscopy (Nikon TE2000-E microscope, Yokugawa CSU-X1 spinning disk, Hamamatsu
ImagEM camera, 60x objective, Metamorph acquisition software). Active contours
were fit to individual MTs using the ImageJ plugin JFilament
\cite{Smith:2010do}, and MT lengths were determined from the contours
using custom MATLAB software. For each taxol concentration, distributions of
MT lengths were fit to a log-normal distribution to find the location
parameter, $\mu$, and the scale parameter, $\sigma$. In each case, the mode microtubule length for the log-normal distribution is given by $$l_{mode} = e^{\mu - \sigma ^2}.$$

\section{Results}
\begin{figure}[h]
\includegraphics[scale=0.9]{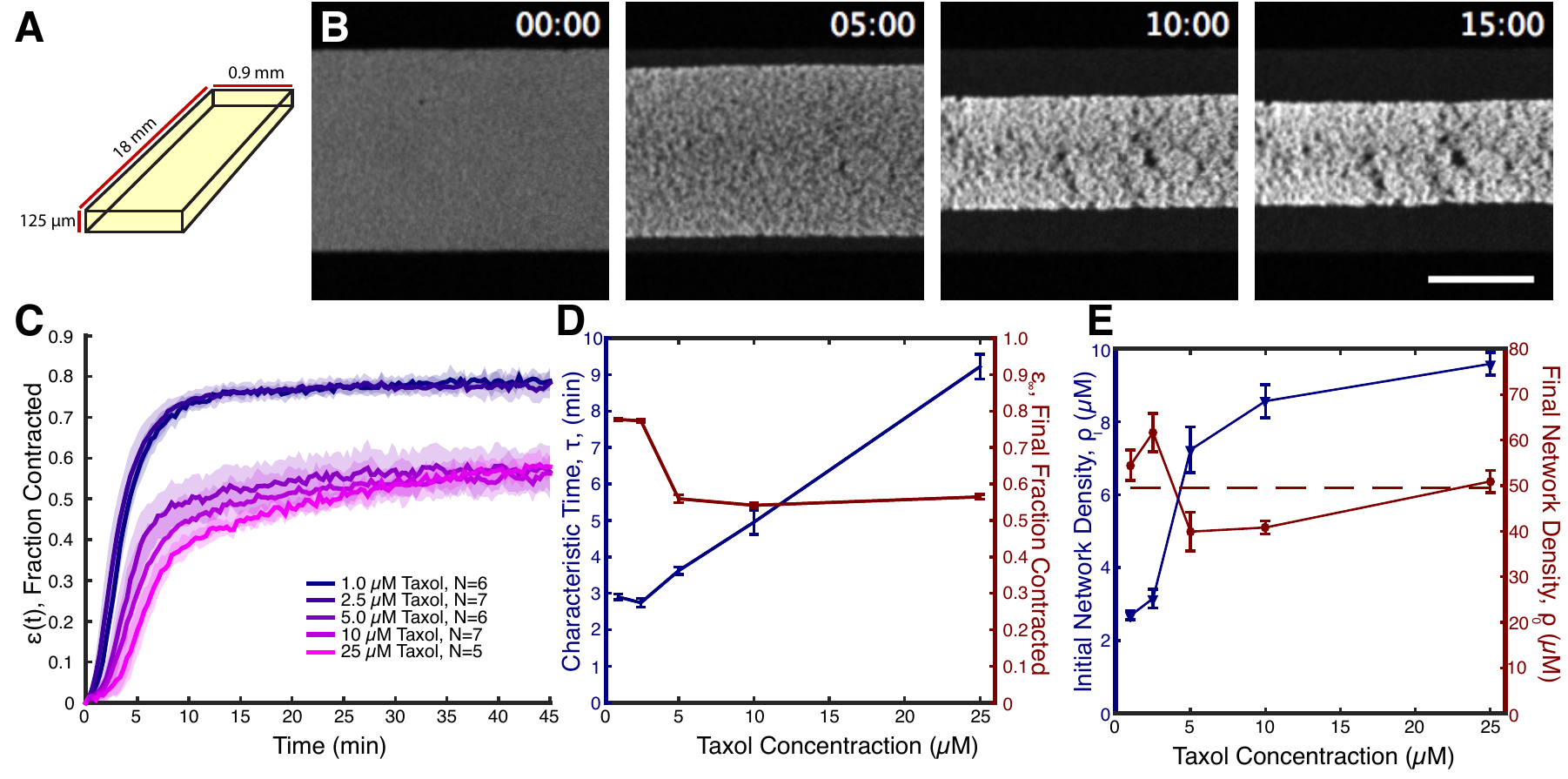}
\caption{{\bf Microtubule networks undergo spontaneous bulk contraction}
(A) Schematic showing characteristic dimensions of the microfluidic chamber. (B)
Time course of MT network contraction when 10 $\mu$M taxol is added
(Scale bar: 500 $\mu$m) (C) Curves showing the fraction contraction,
$\epsilon$(t) as a function of time for varying taxol concentration. Curves are
mean $\pm$ s.e.m. (D) Characteristic time, $\tau$, and final fraction
contraction, $\epsilon_{\infty}$, as a function of taxol concentration. (E)
Initial network density, $\rho_{I}$, (blue line) and final network density,
$\rho_{0}$, (red line) as a function of taxol concentration. The red dashed line
denotes the average value of $\rho_{0}$=49.6 $\mu$M .}
\label{fig:Figure1}
\end{figure}
 
To further investigate the dynamics of contracting MT networks in
\emph{Xenopus} oocyte extracts, we added taxol to the extracts to stabilize
MTs, as previously described \cite{Foster:2015gfa}. Extracts were
loaded into rectangular microfluidic devices (Figure \ref{fig:Figure1}A), sealed
at the inlet and outlet using vacuum grease to prevent evaporation, and imaged
at low magnification (Methods). Within minutes of taxol addition, the MT networks were found to undergo a spontaneous bulk contraction as shown in our previous work \cite{Foster:2015gfa} (Figure \ref{fig:Figure1}B, Movie 1). The size of the network along the width of the channel, $W(t)$, was measured by identifying the pixels with high intensity (belonging to the network) and recorded as a
function of time. Then the fraction contracted $\epsilon(t)$ was calculated, defined as
\begin{equation}
\epsilon(t) = \frac{W_0 - W(t)}{W_0},
\label{eqn:Eq1}
\end{equation}
where $W_0$ is the width of the channel, typically 0.9 mm. This process was
repeated for varying final concentrations of taxol, and the $\epsilon(t)$ curves for
each taxol condition were averaged together to produce master curves for each
condition (Figure \ref{fig:Figure1}C). The $\epsilon(t)$ curves were well fit by a
saturating exponential function,
\begin{equation}
\epsilon(t) = \epsilon_{\infty}[1 - e^{-(\frac{t-T_c}{\tau})}],
\label{eqn:Eq2}
\end{equation}
where $\epsilon_{\infty}$ is the final fraction contracted, $\tau$ is the
characteristic contraction time, and $T_c$ is a lag time between the beginning
of imaging and the beginning of the the contraction process. Equation~\ref{eqn:Eq2}
was fit to the $\epsilon(t)$ curves for each experiment to extract values for the
characteristic time, $\tau$, and the final fraction contracted,
$\epsilon_{\infty}$. Values of $\tau$ and $\epsilon_{\infty}$ were averaged for
each taxol condition (Figure \ref{fig:Figure1}D). The characteristic timescale,
$\tau$, was found to increase with approximate linearity for taxol concentrations
$>2.5 \mu$M, while the final fraction contracted, $\epsilon_{\infty}$, decreased
slightly with increasing taxol concentration. 

We next investigated how changing taxol concentration
influenced both the initial density of the MT network, $\rho_I$, and
the final density of the MT network, $\rho_0$. In systems of purified tubulin, the concentration of tubulin polymerized into
MTs has been shown to increase with increasing taxol concentration
\cite{Kumar:1981tn}. Fluorescence intensity
was used as a proxy for tubulin concentration, and was calibrated using
previously measured values for the total tubulin concentration in \emph{Xenopus}
extracts (Methods). While the initial density of the MT network,
$\rho_I$, monotonically increased with increasing taxol concentration, the final
density of the MT network displayed no obvious trend (Figure
\ref{fig:Figure1}E), and values of the final network density vary from the mean
final density of $\rho_0=49.6 \mu$M by less than 25\%. This is consistent with
previous results arguing that contracting MT networks in \emph{Xenopus}
extracts contract to a preferred final density \cite{Foster:2015gfa}. Also, the increase of 
initial MT density, $\rho_I$, with increasing taxol
concentration appears to saturate for large taxol concentrations.
Considering the three largest taxol concentrations investigated here (5 $\mu$M,
10 $\mu$M, and 25 $\mu$M), the initial MT densities was found to vary
from the mean value by only 14\% (std/mean), far less than the $\approx$
2.5$\times$ increase seen in the characteristic contraction timescale, $\tau$, over this same
range. Thus, $\tau$ does not scale
proportionally to the initial density of MTs, $\rho_I$, and the change
in contraction timescale must come from some other effect of changing taxol
concentration.

 \begin{figure}[t]

\includegraphics[scale=0.9]{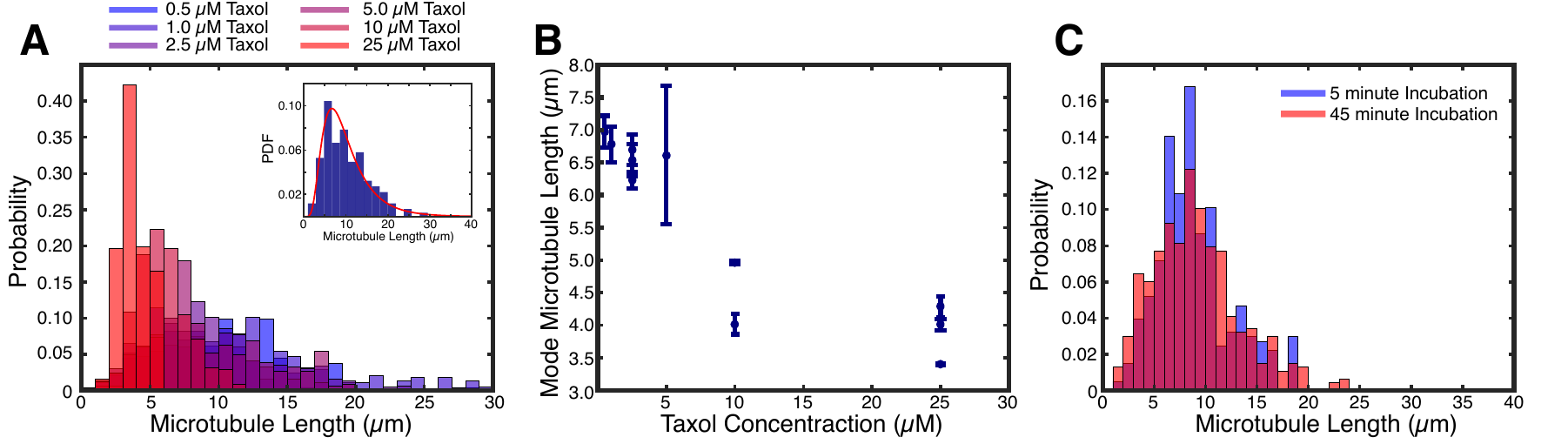}
\caption{{\bf Microtubule length distributions vary with taxol concentration}
(A) Histograms of MT lengths measured for differing taxol
concentrations. Inset: Example distrubution for 2.5 $\mu$M taxol with log-normal
fit (red line). (B) The mode MT length decreases with increasing taxol
concentration. (C) Example histograms for 2.5 $\mu$M taxol, where MT
networks were dissociated and fixed at either 5 minutes or 45 minutes after
taxol addition.}
 \label{fig:Figure2}
 \end{figure}
 
 Previously, it has been shown in clarified \emph{Xenopus} extracts that the size of motor-organized taxol-stabilized MT assemblies, termed ``pineapples", 
 decreased with increasing taxol concentration, presumably due to a decrease in
 the average length of MTs \cite{Mitchison:2013gt}. Thus, we next
 sought to measure the length distribution of MTs in our system for
 varying taxol concentrations. We used a previously described method to
 dissociate and fix MTs \cite{Mitchison:2013gt} (Methods).
 Briefly, MTs were allowed to assemble for 5 minutes after taxol
 addition before being diluted into a dissociation buffer containing 250 mM
 NaCl, which dissociates motor proteins and other MAPs from the MTs.
 After a 2 minute incubation, this mix was diluted again into a fixation buffer
 containing glutaraldehyde. MTs were fixed for 5 minutes, further
 diluted, and thinly spread between a slide and a 22 mm$^2$ coverslip.
 MTs were allowed to adhere for 30 minutes before imaging. An active
 contour was fit to MTs in the images \cite{Smith:2010do}, and the
 length of the active contour was used as a measure for the MT length.
 This process was repeated for each taxol concentration.
 
 Example histograms of MT lengths for each taxol concentration are
 shown in Figure \ref{fig:Figure2}A. Visually, the peak of the distribution
 shifts towards smaller values of MT length for increasing taxol
 concentration. Empirically, we find that the distributions of MT
 lengths are well fit by a log-normal distribution (Figure \ref{fig:Figure2}A,
 Inset). As fitting log-normal distributions to the measured histograms allows a
 more robust estimate of the mode MT length than using a purely
 empirical estimate, we fit the MT length distributions for each taxol
 concentration in order to find $\mu$ and $\sigma$, the two parameters of the
 log-normal distribution, and used these parameters to estimate the mode
 MT length for each condition. The mode MT length was found to
 decrease with increasing taxol concentration, varying by a factor of $\approx$
 1.7 across the taxol conditions measured (Figure \ref{fig:Figure2}B). 
 
One potential concern is that the MT length distribution may vary over
the 45 minute timescale of the contraction experiments. Potentially, this could
be due to a number of factors, including MT depolymerization, severing
of MTs, or other causes. To address this issue, we repeated our length
distribution measurement where MT dissociation and fixation began 45
minutes after taxol addition. We find close agreement between the MT
length distributions measured either 5 minutes or 45 minutes after taxol is
added, indicating that the length distribution is approximately constant over
the timescale of the contraction experiments (Figure \ref{fig:Figure2}C).

\section{Model and Simulation}

Our experimental results suggest that the change in contraction timescale observed for changing taxol concentrations may be due to changes in the MT length distribution. While our earlier continuum active gel theory for MT network
contractions \cite{Foster:2015gfa} captures the global contractile behavior
of the network accurately, understanding the dependence of its parameters on
microscale properties, such as MT length, is beyond the scope of the model. To resolve microscale changes and
study their effects on the network's emergent properties, and to test the dependence of contraction timescale on MT lengths, we turn to simulation.
Our simulation tracks the behavior of a suspension of fixed length MTs actuated
by dynein motors which actively crosslink them and drag them through the fluid. 

\subsection{Model Description}

In our modelling, we represent MTs as rigid spherocylinders that interact
sterically and through motor protein coupling. In principle the fluid in which MTs are suspended couples
their dynamics globally. Since solving the full Stokes problem
numerically is prohibitively expensive we here neglect hydrodynamic many-body
couplings. Instead we approximate the effects of the fluid by a local drag, that
is, each MT acts as if it moved through a quiescent fluid. Note that for a dense
and highly percolated network, one expects long range hydrodynamic interactions
to be screened and hence we argue that they can be safely ignored for our purposes.

Suspensions of passive Brownian spherocylinders are well understood in both theory
and simulation, and our numerical framework (in~\ref{sec:tubuesim}) is based on well-known work \cite{bolhuis_numerical_1997,frenkel_thermodynamic_1988,vroege_phase_1992}. What sets our simulations apart from the previous work is the presence of motor proteins, which have been previously considered with other simulation tools \cite{Nedelec:2007bn}. We model
dynein motors as having a non-moving, fixed crosslinking end and a moving crosslinking end with stochastic binding and unbinding behaviors. The two ends are assumed to be connected by a Hookean spring with a rest length $l_D^0 \simeq \SI{40}{\nano\meter}$, which is  the size of a force-free dynein.  
Collisions amongst dyneins or between dyneins and MTs are ignored for simplicity.
In our model, dyneins have one fixed end, which stays rigidly attached to one MT
throughout the simulation, and one motor end which stochastically binds,
unbinds, and walks along other MTs. When the motor end is free, the dynein is transported
with the MT bound to its fixed end without applying any force to it. Given
the motor's small size, the orientation of such a motor is dominated by Brownian motion,
i.e. is random.
If a free motor head is closer than a
characteristic capture radius $r_{cap}$ to another MT, it binds
with a probability $P_b = \Delta t/\tau_b$, where $\Delta t$ is
time step size, and $\tau_b$ is the characteristic binding time. Should a free end
be close enough to several MTs to bind them, its total binding probability $P_b$
stays the same, and one of the candidate MTs is chosen randomly. Finally, for the
binding location we choose the perpendicular projection the dynein center onto
the target MT. 
Bound motor ends walk towards the minus end of the MT they bind to, and exert a
spring force and torque ($\bF^{motor}=\kappa_D(l_D-l_D^0)\be_D$, $\bT^{motor}=\bbr_D\times\bF^{motor}$),
where $l_D$, $\kappa_D$ are the dynein's current length and its spring constant,
respectively, $\be_D$ is the direction of the Hookean spring force along the direction of dynein, and $\bbr_D$ is a vector pointing from the center of mass of a MT to the point where the dynein binds and applies a Hookean spring force. Consistent with earlier work we impose a force velocity relation
relation \cite{gao_multiscale_2015} and set the velocity
\begin{equation}
    v = v_M \left(1 - \min(F^{motor} / F^{stall},1) \right) ~\mathrm{iff}~ F^{motor} < F^{stall} 
\end{equation}
and $v=0$, otherwise.
Here, $v_M$ and $F^{stall}$ are the motors free velocity and stall forces,
respectively.  

All bound motor ends can stochastically unbind, with a characteristic time 
\begin{equation}
\tau_u = \tau_u^0\exp\left(-F^{motor}/(F^{stall}-F^{motor})\right),
\end{equation}
which becomes $\simeq\exp\left(-F^{motor}/F^{stall}\right)$ for $F^{motor}\ll
F^{stall}$, see \cite{kunwar_mechanical_2011}, and diverges
for $F^{motor}\to F^{stall}$. Here $\tau_u^0$ is the force-free
unbinding time. Finally, dyneins that reach the minus end do not immediately detach, but remain at the minus end, i.e. $v=0$, keeping the same unbinding frequency.

\subsection{Setup and Parameters}

\begin{table}[h]
	\centering
	\caption{The parameters fixed in simulations. \label{tab:parameter} }
	\begin{tabular}{ c c c c} 
		\hline
		Parameter & Explanation & Value & Reference  \\ 
		\hline
		$l_D^0$ & dynein free length & \SI{40}{\nano\meter}  & Estimation \\ 
$F^{stall}$ & dynein stall force& \SI{1}{\pico\newton}& Estimation\cite{kunwar_mechanical_2011}\\ 
$\kappa_D$ & dynein spring constant& $\SI{1}{\pico\newton \per\micro\meter}$ & Estimation\cite{schaffner_biophysical_2006} \\ 
$r_{cap}$ & dynein capture radius& $\SI{80}{\nano\meter}=2l_D^0$ & Assumption\\ 
$v_M$ & dynein max walking velocity& \SI{1}{\micro\meter/s}& Estimation\cite{mckenney_activation_2014} \\ 
$\tau_b$ & dynein binding timescale& ${10^{-3}}{s}$ & Assumed to be short\\ 
$\tau_u^0$ & dynein unbinding timescale& $\SI{10}{s}$& Estimation\cite{rick_biorxiv,mckenney_activation_2014} \\ 
$\eta$ & fluid viscosity & \SI{0.02}{\pico\newton\per\micro\meter\squared\per\second } & Estimation\cite{valentine_mechanical_2005}  \\
$k_BT$ & energy scale at \SI{300}{\kelvin} & $ \SI{4.11e-3}{\pico\newton\micro\meter}$ & \SI{300}{\kelvin} \\
		\hline
	\end{tabular}
\end{table}

For this work we explored how key aspects of the microscopic model change the
emergent behavior of the network. In particular, we varied the number of
dyneins affixed to each MT, $D/M$, and the length distribution of MTs. 

The full model has a large number of parameters. However, many can be fixed from
estimates in the literature. We summarize in Table~\ref{tab:parameter} the parameters that are kept unchanged
throughout the simulations presented here. We also fix the number of MTs in our simulations to be
$\SI{10}{\micro M}$, which is the approximate initial concentration found in experiments with 5 -
\SI{25}{\micro M}  of taxol. Following the experimental results, the MT length distribution is taken to be lognormal
with parameters $(\mu,\sigma)$, with $\sigma$ fixed at $0.5$.
Note that in simulation we use shorter MTs than in the experiments due to
limitations in computational resources, see \ref{sec:mtlength}.
 
In the simulations presented here, each MT carries the same number, $D/M$, of fixed dynein molecules,
attached at random locations on the MT that carries them. Note that this
is different from the assumptions made in earlier work
\cite{Foster:2015gfa}, which has dyneins all fixed to the minus ends of MTs.
This difference is motivated by the numerical observation that networks with
all dyneins at minus ends will precipitate into asters, while networks with
dyneins randomly affixed to MTs are more percolated and globally contract. It's unclear whether this difference is due to an artifact in the current simulations. Further exploring the effects of the localization of motors is a future research direction.

All of our simulations are performed in a simulation box of size
$12\times8\times\SI{8}{\micro\meter}$, with periodic boundary conditions along
the long direction, mimicking the experimental geometry. The initial state is
a random arrangement of MTs, in which all dynein motor
ends are unbound.

\subsection{Simulation Results}

\begin{figure}[t]
	\centering
	\includegraphics[width=\linewidth]{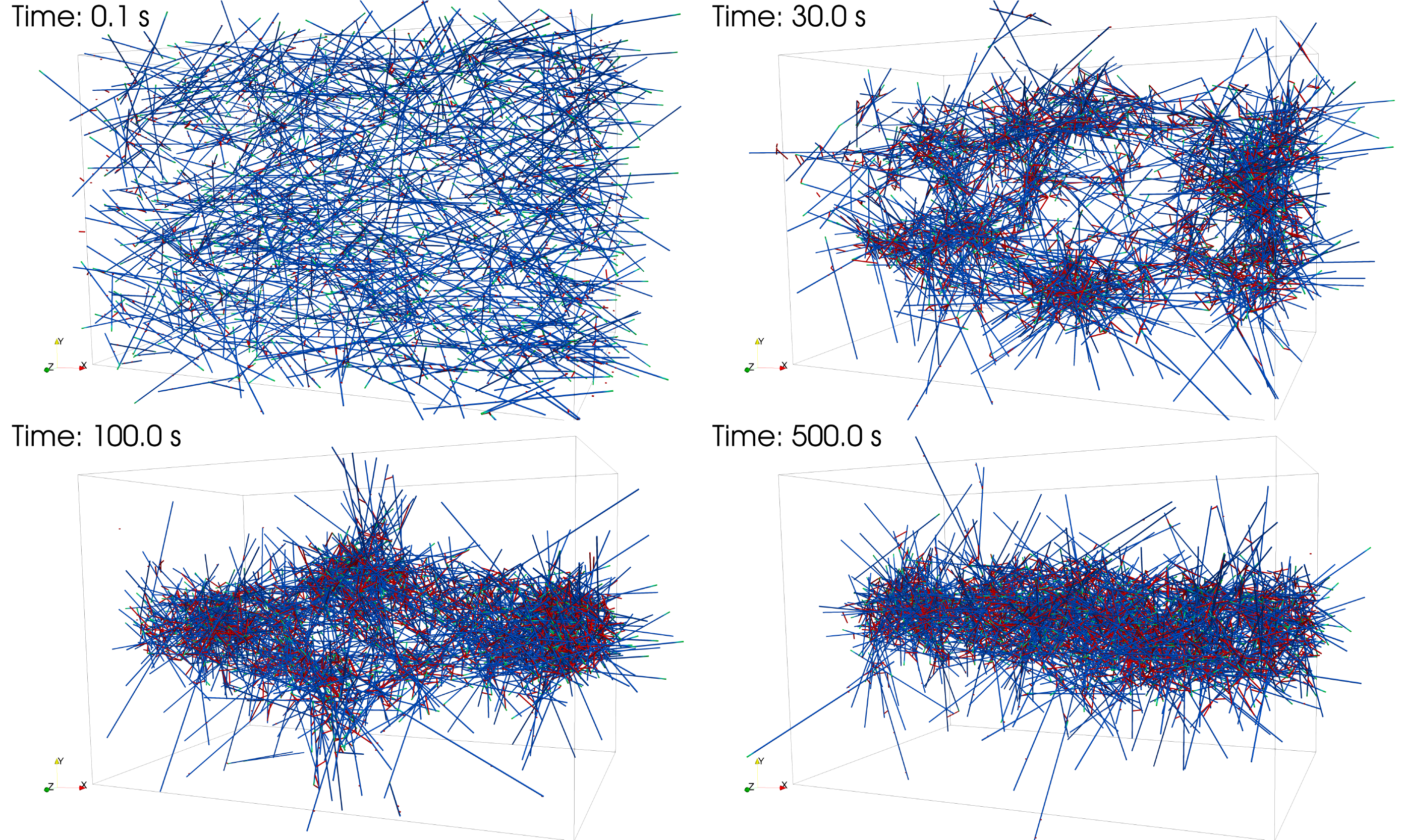}
	
        \caption{\label{fig:simsnap} \textbf{Motor proteins contract the MT network.} The snapshots are taken at different
    time, for the simulation case of $L_{mean}=\SI{2}{\micro\meter}$. In this simulation
    $D/M=2$ which means two dyneins are randomply placed on each MT at
    the beginning of the simulation. The MTs are colored in blue, the
    minus ends are marked by green, and the dyneins are colored in red. The
    simulation box is set to be periodic in the $x$ direction (the horizontal
    direction in the figure), and unbounded in the $y$ and $z$ direction. The
    simulation box size $=12\times8\times\SI{8}{\micro\meter}$, and $1365$
    MTs and $2730$ dyneins are tracked. } \end{figure}

Figure~\ref{fig:simsnap} shows a representative example of the
numerically observed contractions. After an initial phase of $\simeq 30$ s, in
which MTs locally rearrange and cluster, bulk contraction driven by dyneins begins. It takes $\sim 100s$ to create a single, fully percolated network. The system has approached a steady state  at $\simeq 500$ s,
where the network has contracted to a cylindrical ribbon along the periodic direction of
the simulation box.

To compare the simulations to  experiments,
we calculate the fraction contracted 
\begin{equation}
	\epsilon(t) = \frac{R_0-R(t)}{R_0},
        \label{eq:epxilon_simulation}
\end{equation}
from our simulations.  Here, $R(t)$ is the radius of a cylinder aligned with the
periodic direction, and centered around the MT center of mass of the system portion
which contains $50\%$ of the total MT mass.  
By fitting $\epsilon(t)$ to the
saturation exponential, e.g.~Eqn \ref{eqn:Eq2}, we now obtain a contraction timescale
 $\tau$, and a final fraction contracted $\epsilon_\infty$, which can be directly
compared to experiments.
As was done for fitting the experimental results, we ignore the very early stages of contraction
$(\epsilon(t)<0.1)$, since they mostly mirror the initial rearrangement of the
network and not the global contraction dynamics.

Earlier work \cite{Foster:2015gfa} has shown that $\tau$ depends on the
initial system size. Since our numerical system is far smaller than the
experimental one, we do not expect $\tau$ to quantitatively match experiments,
but only to reproduce trends with varying parameters. Using the parameters from
\cite{Foster:2015gfa} for the continuum model, which was done with 2.5 $\mu$M taxol, and a system
size of 8 $\mu$m, we expect a contraction time of $\tau\approx 90$ s, which is close to the time
scales on which our numerical system contracts.
 
\begin{figure}[t]
	\centering
	\includegraphics[width=\linewidth]{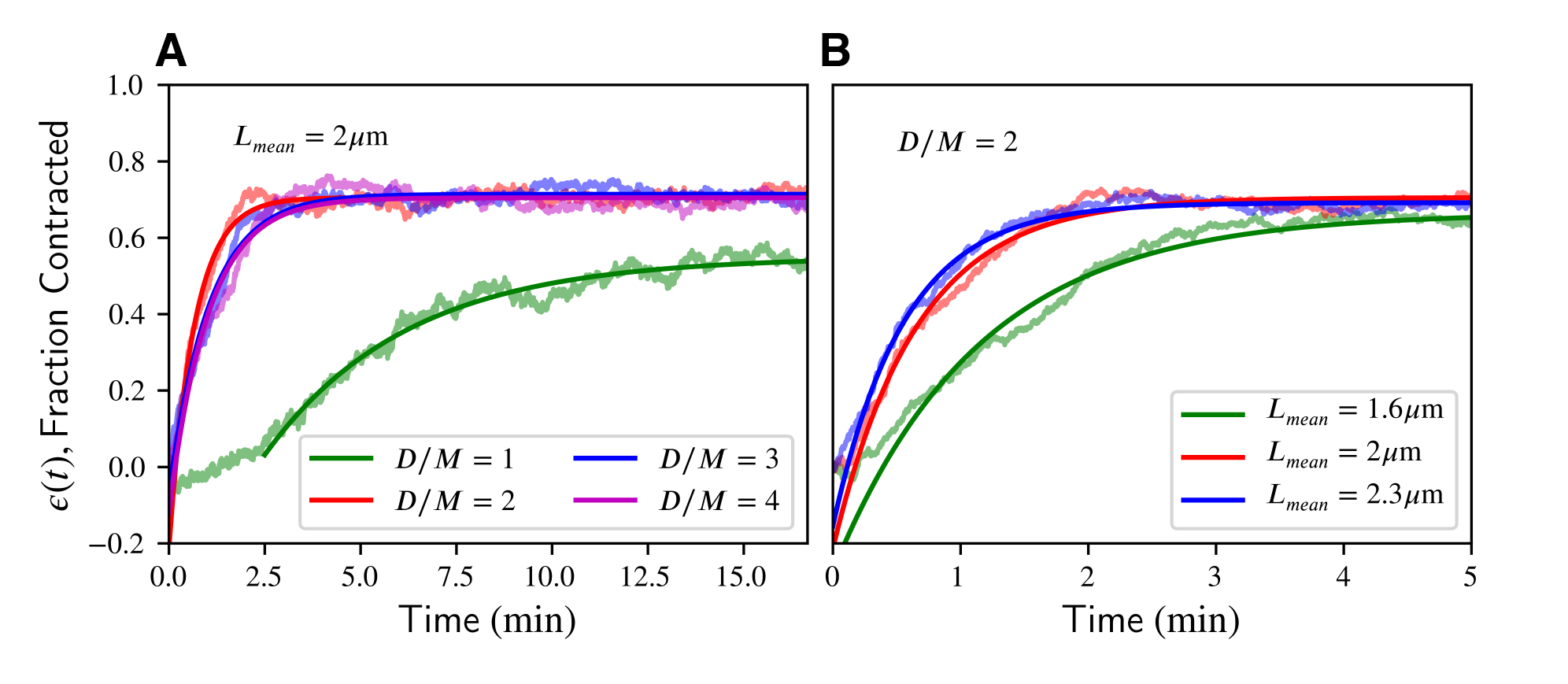}
        \caption{\label{fig:timecontcra} \textbf{The contraction process varies with the dynein number $D/M$ and MT length $L_{mean}$.} Each $\epsilon(t)$ data curve is shown with its
    saturating exponential fit~(\ref{eqn:Eq2}).  The numbers of MTs are
    set at the same value as the $\SI{25}{\micro M}$ taxol case in
    experiments, for all simulations discussed in this paper. More detailed
    setting about the parameters and MT length distribution can be
    found in appendices. } \end{figure}

The final fraction contracted, $\epsilon_{\infty}$, is easier to directly compare between
experiments and simulation, since it does not depend on system size.
Encouragingly, for the case of $D/M=1$, the simulated contraction produces a similar final fraction
contracted of $\epsilon_\infty\approx0.55$, as in experiment, (Figure~\ref{fig:Figure1}D and Figure~\ref{fig:timecontcra}). Note
that here only taxol concentrations of 5, 10, 25 $\mu$M should be directly
compared since the other conditions change the initial density of the system to
values which differ significantly from our simulations.

To further test our numerical model, we investigated how the contraction timescale
varies with the number of motors in the system. As expected from
\cite{Foster:2015gfa}, increasing $D/M$ to $2$ drastically changes the contraction timescale, with only minor
changes in $\epsilon_\infty$ (Figure~\ref{fig:timecontcra}A).  Further
increasing $D/M$ does not significantly change either the timescale $\tau$, or the final fraction $\epsilon_\infty$.  We also tested whether our model can reproduce the changes
of contraction times for different length distributions of MTs,
corresponding to different taxol concentrations, as seen in experiment. Indeed, as
the mean length of MTs is increased, the contractions speed up (Figure~\ref{fig:timecontcra}B, Table \ref{tab:timelength}), yet the final
fraction contracted doesn't change significantly.  This is again consistent with experimental results,
(Figure~\ref{fig:Figure1}). We conclude that our simulations are consistent with key
features of the experiments.

\begin{table}[h]
	\centering
        \caption{The fitted characteristic contraction time $\tau$ for different
        lengths of MTs at $D/M=2$. All simulations start from the same
    network density as the one shown in Figure~\ref{fig:simsnap}.
\label{tab:timelength} } \begin{tabular}{ c c c c c} 
		\hline
		$L_{mean}$ (\si{\micro\meter}) &  1.618 & 2.007 & 2.297 & 2.511  \\ 
		\hline
		$\tau$ (\si{\second}) & 68.6 & 39.4 & 33.4 & 28.8  \\
		\hline
	\end{tabular}
\end{table}

Having gained confidence in our simulation tools, we next inspect important aspects
of the physics of the MT network which are experimentally difficult to access.  We
first ask how the system stress changes during the contraction process.  We
extract the stress from our simulations using the expressions detailed
in~\ref{sec:appendixXF}. In particular, we can distinguish between the stress tensor
contributions from steric  collisions, $\bm{\Sigma}^{col}$, and the active stress tensor generated by motors, $\bm{\Sigma}^{motor}$.
 
\begin{figure}[b]
	\centering
	\includegraphics[width=\linewidth]{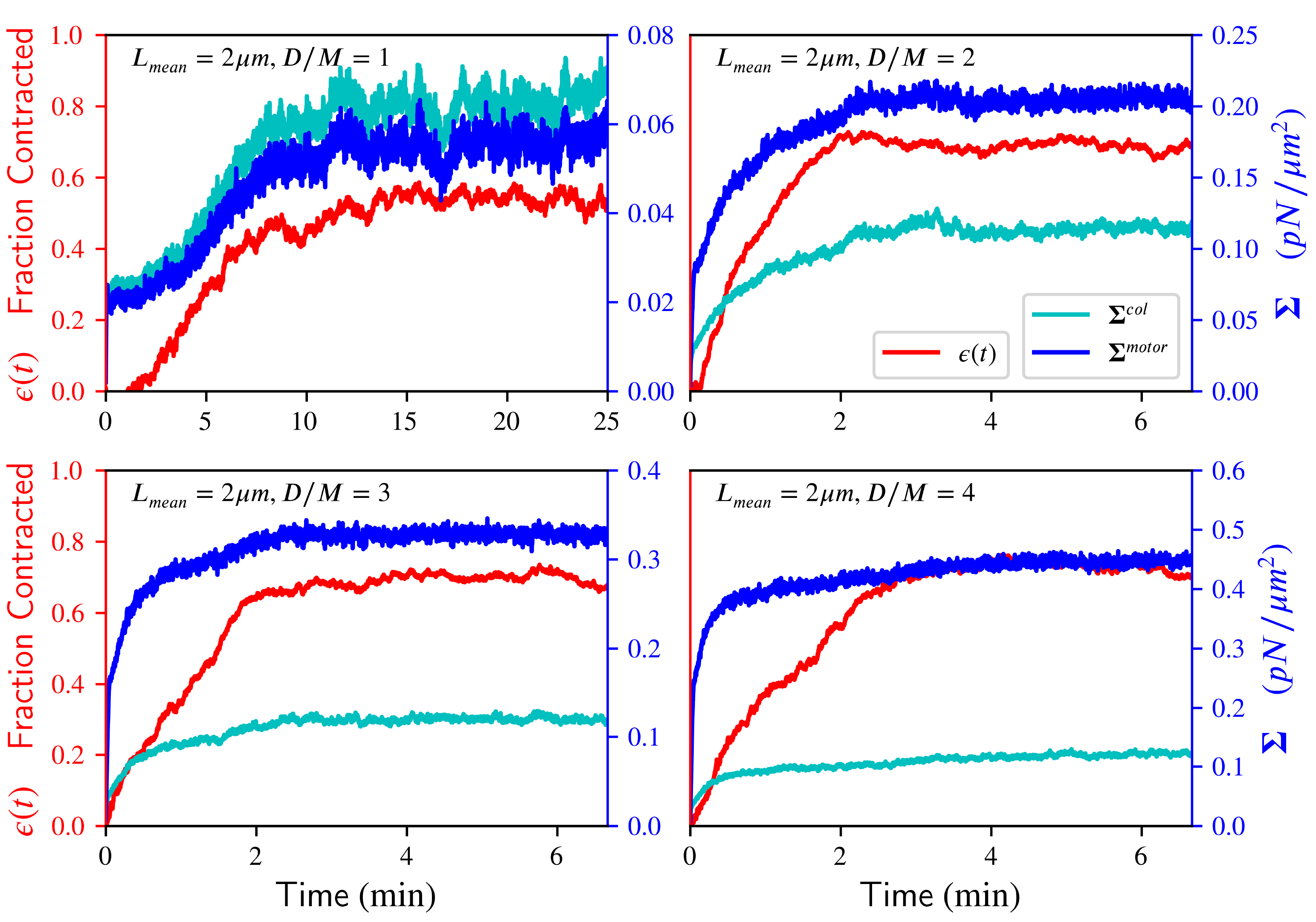}
        \caption{\label{fig:corre} \textbf{The correlation between $\epsilon(t)$ and stress varies with $D/M$.} The fraction contracted is shown in red. The motor stress ${\Sigma}^{motor}$ and the collision stress ${\Sigma}^{col}$ are colored in dark and light blue, respectively. The stresses are calculated with the method described in~\ref{sec:appendixXF}, where the volume $V$ is the volume of the
    entire simulation cell of $12\times8\times\SI{8}{\micro\meter}$. Here the scalar stresses are calculated as the average of diagonal components of the stress tensor: ${\Sigma}^{col} = \frac{1}{3}\left({\Sigma}_{xx}^{col}+{\Sigma}_{yy}^{col}+{\Sigma}_{zz}^{col}\right)$, and ${\Sigma}^{motor}$ is calculated similarly.  }
\end{figure} 

Figure~\ref{fig:corre} shows $\bm{\Sigma}^{col}$,
$\bm{\Sigma}^{motor}$, and $\epsilon(t)$, for $L_{mean}=\SI{2}{\micro\meter}$ and
different numbers of $D/M$, as a function of time. From this data, we make four key
observations. First, we observed that the average stress tensors $\bm{\Sigma}^{col}$ and
$\bm{\Sigma}^{motor}$ are dominated by their diagonal components, and the three diagonal components are within $\sim20\%$ of each other. We report in Figure~\ref{fig:corre} the average of trace of the stress tensor: ${\Sigma}^{col} = \frac{1}{3}\left({\Sigma}_{xx}^{col}+{\Sigma}_{yy}^{col}+{\Sigma}_{zz}^{col}\right)$, and ${\Sigma}^{motor}$ is calculated similarly. Second, we find that at low $D/M$, i.e. 1 or 2, the growth of
$\epsilon(t)$ and of both stress contributions occur on the same timescale. This
breaks down at larger $D/M$ for which the stresses grow and saturate much faster
than $\epsilon(t)$. Third, increasing $D/M$ leads to increasing
$\bm{\Sigma}^{motor}$, while $\bm{\Sigma}^{col}$ rapidly saturates. Fourth, as $\bm{\Sigma}^{col}$
saturates, the overall contraction timescale ceases to decrease (see
also Figure~\ref{fig:timecontcra}).
These observations point to the local structure formation changing significantly
for larger $D/M$, which we further explore.
 
\begin{figure}[t]
	\centering
	\includegraphics[width=\linewidth]{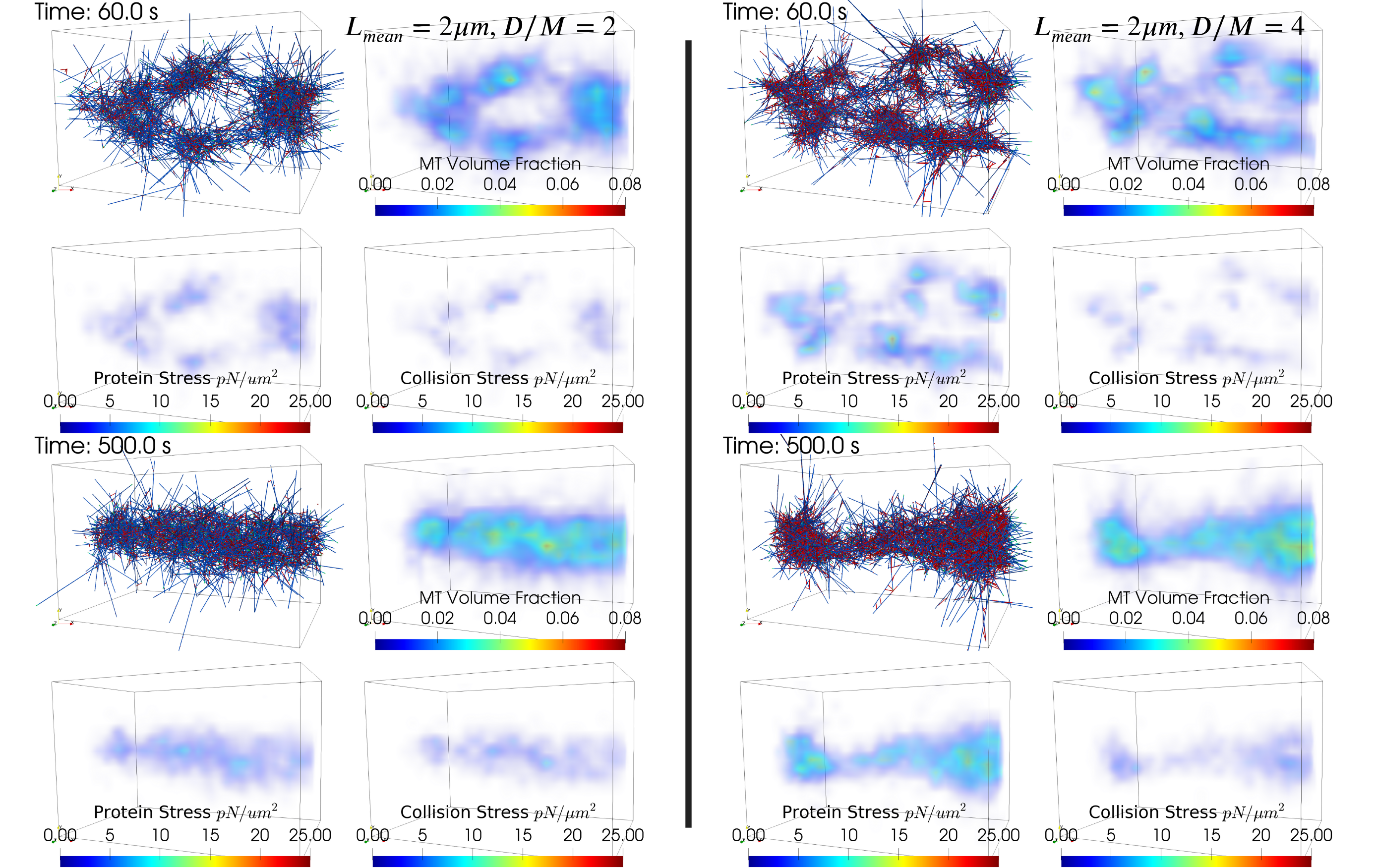}
        \caption{\label{fig:dist}\textbf{The spatial distributions of MT volume fraction and stress vary with the dynein number $D/M$.} Snapshots are taken at $t=\SI{60}{\second}$ and $t=\SI{500}{\second}$ for both cases. On the left:  $L_{mean}=\SI{2}{\micro\meter}$ and $D/M=2$. On the right:
    $L_{mean}=\SI{2}{\micro\meter}$ and $D/M=4$. The spatial distribution is
    calculated by spatial binning with a regular mesh, and the mesh grid size
    $=\SI{0.5}{\micro\meter}$. Here the scalar stresses are calculated as mentioned in Figure~\ref{fig:corre}.}
\end{figure}

Figure~\ref{fig:dist} shows the spatial distribution of MT volume
fraction, motor stress, and collision stress for the $D/M=2$ and $4$ cases.  At
\SI{60}{\second} with $D/M=4$, the MTs are quickly dragged by dyneins
into dense local clusters while the clusters have not contracted much toward the
center. At this stage, $\bm{\Sigma}^{col}$ and $\bm{\Sigma}^{motor}$ have
significantly increased but $\epsilon(t)$ remains low. In contrast, for $D/M=2$
the distibution at $t=60$s is much less clustered.  Consistently for $D/M \leq2$ the growth of $\bm{\Sigma}^{col}$ and $\bm{\Sigma}^{motor}$ is more
synchronized with $\epsilon(t)$. This observation suggests a possible continuum
model to predict the behavior of the entire network based on an
`equation-of-state' relating the local microscopic parameters and the
binding-walking-unbinding cycles of dyneins to the local mechanical stress tensors
$\bm{\Sigma}^{col}$ and $\bm{\Sigma}^{motor}$, as the driving `force' of the
contraction, which will be the subject of future work.

\section{Discussion}
Here we examined how the dynamics of bulk MT network contraction in
\emph{Xenopus} extracts vary with taxol concentration. We find that using
different taxol concentrations, the networks contract to approximately the same
final density, even though the initial density of the networks varies by a
factor of $\approx$ 3.6. While the final density of these networks is constant,
we find that the timescale of the contraction process increases with increasing
taxol concentration. In the high taxol regime, where the initial network density
varies by only $\approx$ 14\%, we find that the timescale increases by a factor
of $\approx$ 2.5, arguing that the timescale is not simply varying
proportionally to the initial network density. The length distributions of the
MTs were measured for each taxol condition, and it was found that the
average MT length decreases with increasing taxol concentration by a
factor of $\approx$ 1.7 across the tested conditions. We postulate that it is this change in MT length that drives the changes in contraction time, and turn to simulation to test this idea.

We built a simulation tool to reveal more microscopic information of the network
contraction process, and found that by placing dyneins randomly on each
MT, we could simulate a contraction process consistent with experiments. We found that the
characteristic contraction timescale, $\tau$, is decreased by increasing only the
MT length, which is consistent with the experimental result that shorter
MTs at higher taxol concentration contract slower. We also
investigated the effect of dynein number on the contraction process. Increasing the number of 
dynein per MT, $D/M$, from $1$ to $2$ sped up the contraction significantly,
but further increasing the dynein number did not affect the contraction. Instead, an increase of $D/M$ generated a local clustering structure
at the early stage of the contraction, and then the local clusters contract.
This process is clearly revealed by the spatial-temporal variations of
collision and motor stresses, and their correlation with the contraction process
$\epsilon(t)$. This indicates that we could possibly build a microscopic
`equation-of-state' to improve our previous coarse-grained
model \cite{Foster:2015gfa} to directly relate the stress terms in the model
with microscopic motor behaviors, which will be the focus of future work.

\section{Acknowledgments}
This work was supported by the Kavli Institute for Bionano Science and Technology at Harvard
University and National Science Foundation grants PHY-1305254,  PHY-0847188, and DMR-0820484 (DJN), and partially supported by National Science Foundation Grants DMR-0820341 (NYU MRSEC), DMS-1463962, and DMS-1620331 (MJS). We thank Bryan Kaye for useful discussions.

\appendix 
\section{Algorithm for MT network simulation}
\label{sec:tubuesim}
The simulation program tracks the motion of MTs driven by motor
proteins by modeling MTs as rigid spherocylinders and motor proteins as Hookean springs with crosslinking ends. At this micron-sized scale the motion of objects in fluid is overdamped, and MTs are tracked
with the following simple explicit Euler time stepping, where $\bX_i$ is the center of mass and $\bq_i$ is the unit orientation vector of MT $i$:
\begin{eqnarray}
\Delta\bX_i  &= \bM_{i} \cdot \left( \bF_i^{motor} + \bF_i^{col} \right) \Delta t + \Delta\bX_i^B \label{eq:simtrans},\\
\Delta\bq_i  &= \frac{1}{\gamma_i^R} \left(\bT_i^{motor} + \bT_i^{col}\right) \times \bq_i \Delta t + \Delta\bq_i^B. \label{eq:simrot}
\end{eqnarray} 
The forces $\bF^{motor}$ and $\bF^{col}$ stem from motor proteins and
collisions between MTs, respectively, with $\bT^{motor}$ and
$\bT^{col}$ being the corresponding torques. Finally $\Delta\bX^B$ and
$\Delta\bq^B$ are the Brownian contributions to the motion of MTs.

The mobility matrix $\bM_{i}$ describes the effect of hydrodynamic drag. In
principal all MTs are fully coupled through the Stokes
equation, but in this work we ignore this coupling. Here, the $\bM_i$ of each MT $i$ is approximated as if
each MT is moving individually by itself in unbounded fluid. $\bM_i$
takes a $3\times3$ symmetric matrix form because we modeled each MT
as an axisymmetric spherocylinder:
\begin{equation}
	\bM_i^{-1} = \gamma_i^\parallel \bq_i\bq_i + \gamma_i^\perp (\bI - \bq_i\bq_i),
\end{equation}
where $\bI$ is the $3\times3$ identity matrix. $\gamma_i^\parallel$ and
$\gamma_i^\perp$ are the translational drag coefficients for motions parallel
and perpendicular to the axis of the MT $i$, depending on the fluid
viscosity $\eta$, the MT length $L_i$, and the diameter $D_i$. Drags
are solvable from the slender-body theory\cite{gotz_analysis_2000}:
\begin{eqnarray}
	\gamma_i^\perp = \frac{8\pi\eta L_i}{b_i+2},\quad \gamma_i^\parallel = \frac{8\pi\eta L_i}{2b_i}, \quad \gamma_i^r =\frac{2\pi\eta L_i^3}{3(b_i+2)}\label{eq:dragslender},
\end{eqnarray}
where $b_i=-\left(1+2\ln (D_i/L_i)\right)$ is a geometric parameter. For
MTs we have $L_i\gg D_i$, and $b_i\approx2\ln(L_i/D_i)$, and so we have
$\gamma_i^\perp \approx 2\gamma_i^\parallel$. In simulations, we used the same
diameter $D$ for all MTs. 

At each timestep the Brownian displacement and rotation $\Delta \bX_i^B$ and $\Delta\bq_i^B$ are
generated as Gaussian random vectors. The means are zero, while
the variances are given by the local fluctuation-dissipation relation:
\begin{eqnarray}
\AVE{\Delta \bX_i^B\Delta \bX_i^B} &= 2k_B T \bM_i\Delta t, \\
\AVE{\Delta \bq_i^B\Delta \bq_i^B} &= 2k_B T \left(\bI-\bq_i\bq_i\right) \Delta t/{\gamma_i^r}.
\end{eqnarray}
We follow the method of Tao et al.\cite{tao_brownian_2005} to generate these two
random vectors.

For Eq.~\ref{eq:simtrans}, the motor force and torque are described in the main text. The collision force $\bF_i^{col}$ and torque $\bT_i^{col}$ are calculated in the simulation as detailed in~\ref{sec:ftcol}.

\section{Collision between microtubules} 
\label{sec:ftcol}
When two MTs $i$ and $j$ are close to each other, the shortest distance
$d_{ij}$ between their axes is calculated. If $d_{ij}$ is smaller than the microtubule diameter $D$ then a collision force
$\bF^{col}$ is calculated with a WCA-type repulsive force to prevent them from
overlapping and crossing each other. The original WCA potential is stiff and
poses a severe limit on the maximum timestep $\Delta t$ in simulations. We used a softened WCA potential to enable larger timesteps:
\begin{eqnarray}
d\geq\beta D: \quad	F^{col}(d) &= -24 \left(\frac{k_BT}{D}\right)\left[\frac{1}{\left(\alpha+d/D\right)^{7}}-\frac{2}{\left(\alpha+d/D\right)^{13}}\right], \\
d<\beta D: \quad F_L^{col}(d) &=  F^{col}(\beta D) + \dpone{F^{col}}{d}|_{d=\beta D} (d-\beta D) .
\end{eqnarray}
Where $F^{col}(d)$ is a shifted WCA potential with a shift parameter $\alpha$.
For $\alpha=0$, $F^{col}(d)$ becomes the usual WCA repulsive force.
$F_L^{col}(d)$ is a linearized continuation of $F^{col}(d)$ for $d<\beta D$, to
prevent the blow-up of collision force at small $d$. In simulations we fix
$\alpha=0.1$ and $\beta=0.95$, to preserve the repulsion between MTs,
but enable the simulations to complete within reasonable simulation time. Also
due to the use of a soft repulsive force instead of an accurate rigid
interaction, the effective diameter of MTs are estimated to be about
$\sim 80\%$ of the specified diameter $D$ \cite{tao_brownian_2005}. Therefore in
simulations $D$ is set to $\SI{30}{\nano\meter}$ to reproduce the true diameter
\SI{24}{\nano\meter} of MTs.

\section{The calculation of stress and its spatial distribution}
\label{sec:appendixXF}
The stress can be calculated by the binding force and collision force in Eq.~(\ref{eq:simtrans}). Since both contributions are pairwise between MTs, and satisfy Newton's third law, we can calculate the stress tensor pair-by-pair with the virial theorem. Also, because the MTs are thin-and-long slender cylinders, the calculation can be further simplified \cite{snook_normal_2014}.

For two MTs $i$ and $j$, their contribution to the total system stress is a $3\times3$ tensor formed by the outer product:
\begin{equation}
\bm{\sigma}^{(i,j)}	= - \left(\bx^{(i)}	-\bx^{(j)} 	\right)\bff^{(i,j)}	 ,
\end{equation}
where $\bbr^{(ij)}=\bx^{(i)}-\bx^{(j)}$, is the vector pointing from the forcing location $\bx^{(j)}$ on MT $j$ to $\bx^{(i)}$ on MT $i$. $\bff^{(i,j)}$ is the force from $j$ to $i$. For the collision force,  $\bx^{(i)}$ and $\bx^{(j)}$ are the two collision points on each MT. For the dynein binding force, $\bx^{(i)}$ and $\bx^{(j)}$ are the two binding locations (two ends) of one dynein on the two MTs. If more than one dyneins bind $i$ and $j$, the contribution from each dynein is calculated and added independtly. 

Each $\bm{\sigma}^{(i,j)}$ contributes to the total system stress tensor at the spatial location $\left(\bx^{(i)}+\bx^{(j)}\right)/2$, which is the center of the two forcing location. For a certain volume $V$ in space, the average stress tensor in this volume is a simple arithmetic average of all $\bm{\sigma}_{ij}$ in this volume:   
\begin{equation}
	\bm{\Sigma}_V^{col} = -\frac{1}{V} \sum_{\in V} \bm{\sigma}^{(i,j),col},\quad\bm{\Sigma}_V^{motor} = -\frac{1}{V} \sum_{\in V}  \bm{\sigma}^{(i,j),motor}.
\end{equation}

\section{The length distribution of microtubules}
\label{sec:mtlength}
The length distribution of MTs is taken to be given by a lognormal distribution with parameters $\mu$ and $\sigma$:
\begin{equation}
	P(L) = \frac{e^{-\frac{(\log (L)-\mu )^2}{2 \sigma ^2}}}{\sqrt{2 \pi } L \sigma }, \quad L_{mean} = e^{\mu+\sigma^2/2}, \quad L_{mode} = e^{\mu-\sigma^2}.
\end{equation}

In the simulations, to detect the binding and collision events a nearest neighbor search procedure is performed during each timestep, and long MTs significantly slow down its efficiency. Also, due to the periodic boundary condition used in simulations, a MT with length of the simulation box size may interact with its own periodic image, and generates some unphysical results. Therefore in simulations the lognormal distribution is clipped at $L_{max}$, and the mean length of the MTs changes:
\begin{equation}
L_{mean} = \frac{e^{\mu +\frac{\sigma ^2}{2}} \mathrm{Erfc}\left(\frac{\mu -\log (L_{max})+\sigma ^2}{\sqrt{2} \sigma }\right)}{\mathrm{Erfc}\left(\frac{\mu -\log (L_{max})}{\sqrt{2} \sigma }\right)} .
\end{equation}
While the mode length does not change with $L_{max}$.

In simulations we used $L_{max}=\SI{4}{\micro\meter}$, which is half of the width of a simulation box with size $12\times8\times8\si{\micro\meter}$. We also fixed $\sigma=0.5$ to match the experiment data, and the actual simulation settings are detailed in Table~\ref{tab:length}.
\begin{table}[h]
	\centering
		\caption{The settings of MT length in simulations. 	\label{tab:length} }
	\begin{tabular}{ c c c c c c} 
		\hline
		$e^\mu$ (\si{\micro\meter}) & 1.0 &  1.5 & 2.0 & 2.5 & 3.0 \\ 
		\hline
		$L_{mean}$ (\si{\micro\meter}) & 1.123 & 1.618 & 2.007 & 2.297 & 2.511  \\ 
		$L_{mode}$ (\si{\micro\meter}) & 0.779 & 1.168 & 1.558 & 1.947 & 2.336  \\ 
		\hline
	\end{tabular}
\end{table}


\end{document}